\documentclass[
reprint,
superscriptaddress,
amsmath,amssymb,
aps,
]{revtex4-2}

\usepackage{graphicx}
\usepackage{dcolumn}
\usepackage{gensymb}
\usepackage{color}
\usepackage{hyperref}
\hypersetup{colorlinks,
			linkcolor={blue!75!black!80!yellow},
			citecolor={blue!75!black!80!yellow},
			urlcolor={blue!75!black!80!yellow},
			pdfstartview=FitH}
\usepackage{graphicx}
\usepackage{float}
\usepackage{outlines}
\usepackage{enumitem}
\usepackage{upgreek}
\newcommand{\beginsupplement}{
    \setcounter{table}{0}
    \renewcommand{\thetable}{S\arabic{table}}
    \setcounter{figure}{0}
    \renewcommand{\thefigure}{S\arabic{figure}}
}
\usepackage{xcolor}

\usepackage{color,soul} 

\begin{document}

\title{Spectral control of nonclassical light using an integrated thin-film lithium niobate modulator}

\author{Di Zhu,$^{1,2,\dagger,*}$
Changchen Chen,$^{3,\dagger}$
Mengjie Yu,$^{1,\dagger}$ 
Linbo Shao,$^{1}$
Yaowen Hu,$^{1}$
C. J. Xin,$^{1}$
Matthew Yeh,$^{1}$
Soumya Ghosh,$^{1}$
Lingyan He,$^{4}$
Christian Reimer,$^{4}$
Neil Sinclair,$^{1}$
Franco N. C. Wong,$^{3}$
Mian Zhang,$^{4}$
Marko Lon\v{c}ar$^{1,*}$\\
\small
\textit{$^1$John A. Paulson School of Engineering and Applied Sciences,\\ Harvard University, Cambridge, MA 02138, USA\\
$^2$Institute of Materials Research and Engineering, Agency for Science, Technology and Research (A*STAR), Singapore 138634, Singapore\\
$^3$ Research Laboratory of Electronics, Massachusetts Institute of Technology, Cambridge, MA 02139, USA\\
$^4$ HyperLight Corporation, 501 Massachusetts Ave, Cambridge, MA 02139, USA\\
$^\dagger$These authors contributed equally}\\
$^*$\href{mailto:zhu_di@imre.a-star.edu.sg}{zhu\_di@imre.a-star.edu.sg}; \href{mailto:loncar@seas.harvard.edu}{loncar@seas.harvard.edu}
}

\date{Dec 17, 2021}
\begin{abstract}
Manipulating the frequency and bandwidth of nonclassical light is essential for implementing frequency-encoded/multiplexed quantum computation, communication, and networking protocols~\cite{Lukens2017, Kues2017, Sinclair2014, Kimble2008TheInternet}, and for bridging spectral mismatch among various quantum systems~\cite{Aharonovich2016,Lvovsky2009OpticalMemory,Awschalom2018QuantumSpins}. However, quantum spectral control requires a strong nonlinearity mediated by light, microwave, or acoustics~\cite{Karpinski2017,Wright2017,Lavoie2013,Fan2016,Raymer2012ManipulatingPhotons}, which is challenging to realize with high efficiency, low noise, and on an integrated chip. Here, we demonstrate both frequency shifting and bandwidth compression of nonclassical light using an integrated thin-film lithium niobate (TFLN) phase modulator. We achieve record-high electro-optic frequency shearing of telecom single photons over terahertz range ($\pm$ 641 GHz or $\pm$ 5.2 nm), enabling high visibility quantum interference between frequency-nondegenerate photon pairs. We further operate the modulator as a time lens and demonstrate over eighteen-fold (6.55\,nm to 0.35\,nm) bandwidth compression of single photons. Our results showcase the viability and promise of on-chip quantum spectral control for scalable photonic quantum information processing.  
\end{abstract}

\maketitle

Optical photons are ideal carriers of quantum information, as exemplified by their widespread and indispensable use in quantum information science. Compared to commonly used degrees of freedom such as polarization~\cite{OBrien2003} and path~\cite{Kok2007}, encoding and processing quantum information in the frequency domain of a photon promises massive channel capacity and operation parallelism in a single waveguide~\cite{Lukens2017, Kues2017}. The ability to manipulate photon spectra is a prerequisite to implement frequency-domain protocols and schemes, such as spectral linear optical quantum computing~\cite{Lukens2017}, as well as frequency-multiplexed quantum repeaters~\cite{Sinclair2014} and quasi-deterministic single-photon sources~\cite{GrimauPuigibert2017,Joshi2018}. In addition, different sources of single photons, produced by heralding photon pairs~\cite{Kwiat1995NewPairs,Yan2015GenerationWaveguide} or from solid-state emitters~\cite{Aharonovich2016}, vary largely in frequency and bandwidth. Spectral control of single photons is therefore crucial for matching frequency/bandwidth differences and inhomogeneities among these photon sources, and for interfacing them with spectrally mismatched quantum memories~\cite{Kimble2008TheInternet, Lvovsky2009OpticalMemory}. 

Quantum spectral control, however, has proven difficult as it requires altering photon energies without introducing loss or noise. The most common approaches rely on optical nonlinearities~\cite{Raymer2012ManipulatingPhotons}, such as four-wave mixing Bragg scattering~\cite{McKinstrie2005,Singh2019,Li2016EfficientNanophotonics}, sum-/difference-frequency generation~\cite{Lavoie2013,Kumar1990QuantumConversion, Zaske2012,Allgaier2017HighlyLight}, and cross-phase modulation~\cite{Matsuda2016DeterministicModulation}. These processes can achieve large frequency shifts but involve strong optical pumps, which are prone to generating noise photons and require stringent pump filtering. Optomechanical frequency shifters have also been explored, featuring small device footprint and on-chip integrability~\cite{Fan2016,Fan2019SpectrotemporalNanomechanics}. However, they require suspended waveguides and high-Q mechanical resonances, which can only operate over a very narrow radio-frequency (RF) bandwidth. Alternatively, electro-optic (EO) phase modulation allows deterministic spectral-temporal control by directly interfacing microwave and optical fields~\cite{Wright2017, Karpinski2017}. Unfortunately, most integrated photonic platforms, such as silicon and silicon nitride, do not offer low-loss, high-bandwidth, and efficient electro-optic phase modulation. Previous demonstrations of EO quantum spectral control have therefore been limited to discrete, bulk modulators~\cite{Karpinski2017,Wright2017,GrimauPuigibert2017,Lu2019}, which hinders their scalability towards complex, large-scale quantum systems. 

In this work, we report on-chip spectral control of nonclassical light using a thin-film lithium niobate (TFLN) modulator. As an emerging integrated photonic platform, TFLN offers a low loss and wide transparency window, as well as large EO, piezo-electric, and $\chi^{(2)}$-nonlinear coefficients~\cite{Zhu2021}. Notably, TFLN EO modulators have demonstrated significant advantages over their traditional bulk counterparts in terms of half-wave voltage ($V_\pi$), bandwidth, footprint, and integrability~\cite{Zhang2021}. Here, we use a specially designed double-pass TFLN phase modulator~\cite{Yu2021} to demonstrate electro-optic spectral control of heralded single-photon pulses at telecom wavelengths. Due to the low voltage requirement of the modulator, it becomes possible to drive multiple $V_\pi$ with readily available RF sources, which in turn enabled us to achieve terahertz-scale ($\pm$ 641 GHz or $\pm$ 5.2 nm) frequency shifts--the largest EO shifting demonstrated to date--along with an 18-fold bandwidth compression through spectral shearing and time lensing, respectively. Our results show a substantial performance breakthrough, and hold great promise for future large-scale, multi-functional integration with other essential classical and quantum components that have already been developed on the TFLN platform~\cite{Zhu2021,Saravi2021}.

\begin{figure}
    \centering
    \includegraphics[width = 3in]{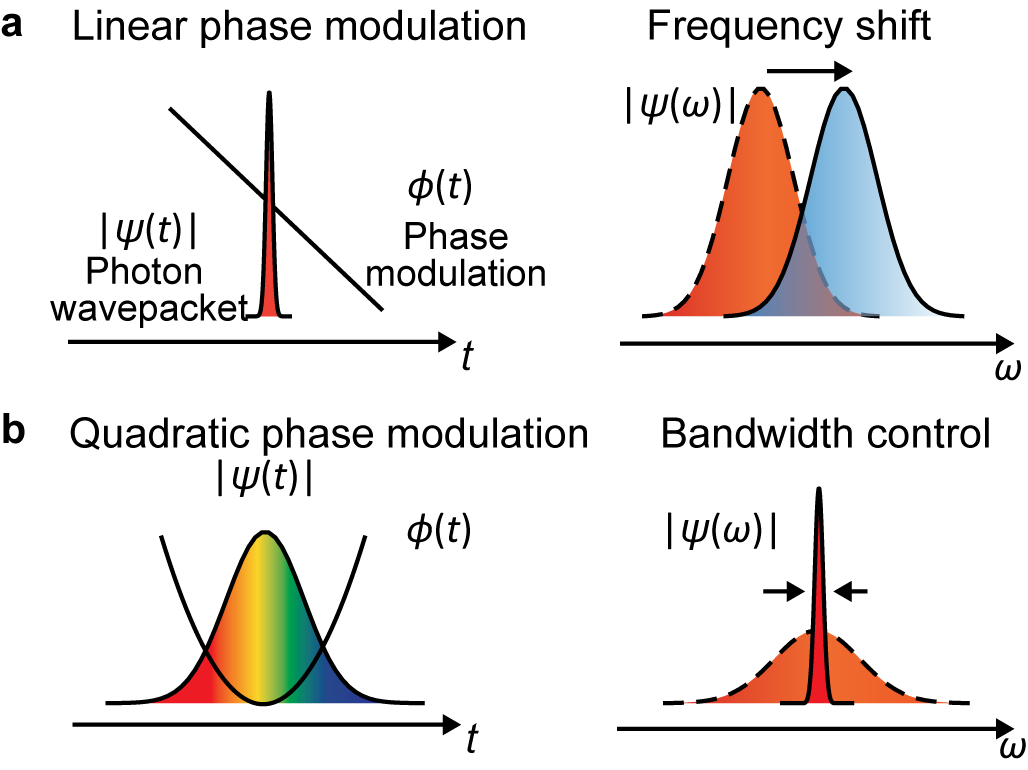}
    \caption{\textbf{Frequency and bandwidth control of light through temporal phase modulation.} \textbf{a}, A linear phase modulation, referred to as spectral shearing, results in a Doppler-like frequency shift. \textbf{b}, A quadratic phase modulation, known as time lens, applied to a properly dispersed optical pulse can lead to bandwidth compression/expansion. }
    \label{fig:fig1}
\end{figure}

\begin{figure}
    \centering
    \includegraphics[width = 3in]{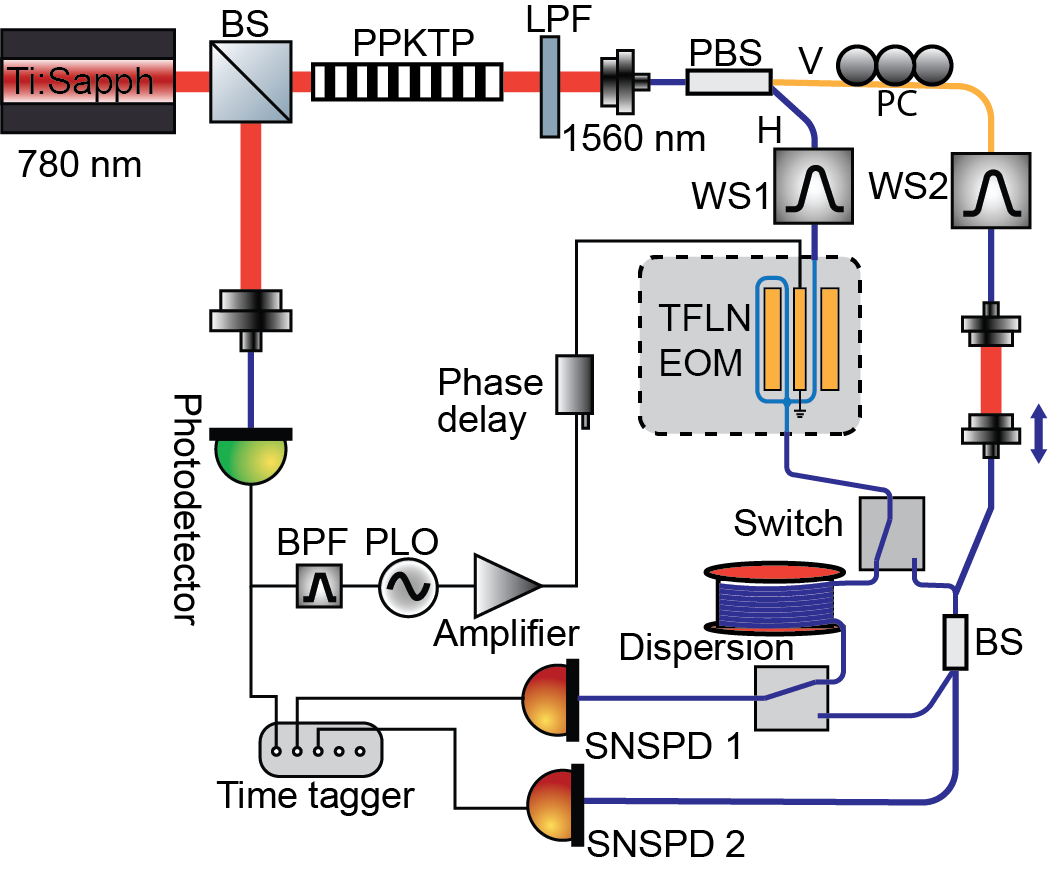}
    \caption{\textbf{Experimental set-up.} Orthogonally polarized photon pairs were generated by pumping a type-II phase-matched periodically poled potassium titanyl phosphate (PPKTP) crystal with a mode-locked Ti:sapphire laser. After being separated by a fiber polarization beamsplitter (PBS), the signal photons (horizontally polarized, H) went through a double-pass TFLN modulator (EOM) with $V_\pi$ as low as 2.3–2.8 V in 10–40 GHz range. The output either passed through a 20 km fiber (labeled Dispersion) followed by superconducting nanowire single-photon detectors (SNSPDs) for dispersion-based spectrum measurement or a 50:50 beam splitter (BS) for quantum interference with the idler photon (vertically polarized, V), which passed through a polarization controller (PC) and a tunable delay. The RF drive of the modulator was phase-locked to the pump laser with a tunable phase delay. Waveshapers (WS) were used to apply spectral filtering or spectral phase; LPF: long-pass filter; BFP: bandpass filter; PLO: phase-locked oscillator.}
    \label{fig:fig2}
\end{figure}

Because of the fundamental Fourier relation between frequency and time, spectral control of light can be realized by temporal phase modulation (Fig.~\ref{fig:fig1}). A linear temporal phase $\phi(t)=-Kt$ applied to a photon wavepacket will lead to a Fourier shift that converts the photon frequency from $\omega$ to $\omega+K$, where $t$ is time in the moving frame of the wavepacket (Fig.~\ref{fig:fig1}a). This Doppler-like frequency shift is often called spectral shearing~\cite{Wright2017,Duguay2004OPTICALPULSES}. On the other hand, a quadratic phase modulation, $\phi(t) = Bt^2/2$, can enable spectral compression or expansion of a photon wavepacket with a properly applied quadratic spectral phase, $\varphi(\omega) = \Phi(\omega-\omega_0)^2/2$ where $B$ is the chirp factor, $\Phi$ is the group delay dispersion (GDD), and $\omega_0$ is the center frequency~\cite{Karpinski2017} (Fig.~\ref{fig:fig1}b). This method is widely used for spectral-temporal control of ultrafast pulses and can be explained by exploiting the space-time duality between paraxial diffraction of a spatially confined optical beam and spectral dispersion of a temporally confined optical pulse~\cite{Kolner1994,Salem2013ApplicationProcessing}. Specifically, mapping to a spatial imaging system, the quadratic temporal phase resembles a lens in time domain (referred to as time lens), and the spectral dispersion is analogous to spatial diffraction in free-space propagation. As a result, for instance, a broadband optical pulse can be ``collimated'' to narrowband when it is placed at the “focal point” of the time lens ($\Phi = 1/B$). The shearing rate and lens curvature, given a certain RF drive power, are ultimately determined by the half-wave voltage ($V_\pi$) and operating frequency of the phase modulator. It is worth mentioning that both spectral shearing and time lensing rely on pure phase modulation and are fundamentally unitary and deterministic.

\begin{figure*}[tbh]
    \centering
    \includegraphics[width = 4.5in]{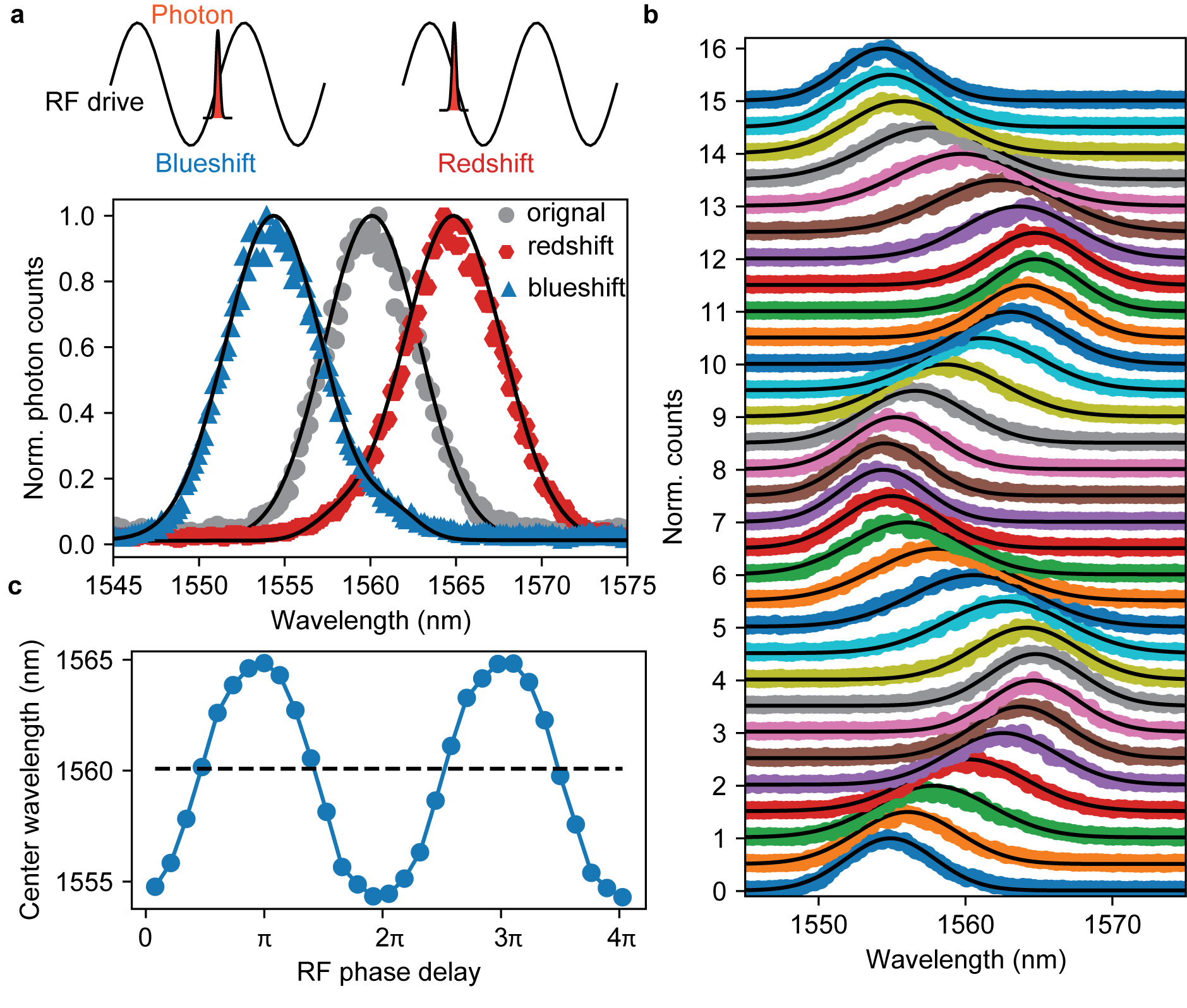}
    \caption{\textbf{Single-photon spectral shearing.} \textbf{a}, Measured single-photon spectra when the photon was locked at the rising and falling slopes of the sinusoidal RF drive. \textbf{b}, Waterfall plot of the measured single-photon spectra with increasing RF phase delay ($\sim0.13\pi$ step), showing that the center frequency can be tuned continuously. The spectra were normalized to their maximum value to account for coupling drift during measurement and displaced vertically for ease of visualization. Due to timing jitter between the RF drive and optical pulse, we observed a spectral broadening of photon bandwidth, especially near zero-shift positions. Black solid lines are Gaussian fittings. \textbf{c}, Extracted frequency shift as a function of RF phase delay, showing maximum shift of $\pm$ 5.2 nm ($\pm$ 641 GHz). The input photon was at 1560 nm with a FWHM bandwidth of 6.5 nm. The modulator was driven at 27.5 GHz with an amplitude of 8.1 $V_\pi$ ($V_\pi\approx2.5$ V). }
    \label{fig:fig3}
\end{figure*}

In our experiment, we generated pulsed photon pairs at ~1560 nm through spontaneous parametric down-conversion (SPDC)~\cite{Chen2017}(Fig.~\ref{fig:fig2}). The orthogonally polarized signal and idler photons were separated using a polarization beamsplitter, and the signal photon was sent to an integrated TFLN phase modulator. Notably, the TFLN modulator features a novel double-pass design, where the optical waveguide passes through the velocity-matched coplanar waveguide electrode twice thus doubling the interaction length between RF and optical fields, resulting in enhanced modulation efficiency (see supplementary Fig.~\ref{fig:fig_s1}). It had a measured $V_\pi$ between 2.3-2.8 V at phase-matched frequencies from 10 to 40 GHz, which is significantly lower than commercially available discrete bulk LN modulators (typically $>$7 V at 30 GHz for telecom wavelength; see, for example, Thorlabs LN27S-FC) as well as previously demonstrated TFLN phase modulators~\cite{Ren2019}. The sinusoidal RF drive for the modulator was phase-locked to the pump laser, and the phase relative to the optical pulses was controlled using a tunable RF delay. Since the optical pulse ($<$1 ps) was much shorter than the RF period (tens of ps), the phase modulation experienced by the optical pulse can be reliably tuned from nearly linear (photon arrival synchronized to the rising/falling edges of the sinusoidal drive) to predominantly quadratic (photon arrival synchronized to the valley/peak of the RF drive) by controlling the relative RF phase. After the TFLN modulator, the photons were either sent to time-based spectrum measurement~\cite{Chen2017,Avenhaus2009} or a beam splitter for quantum interference (see Fig.~\ref{fig:fig2}).

We first demonstrated spectral shearing by driving the modulator with $\sim$8.1 $V_\pi$ at 27.5 GHz (i.e., $\sim$4 W on-chip RF power). By synchronizing the photon to the steepest rising (falling) slope of the RF drive, we obtained redshifted (blueshifted) versions of the original spectra (Fig.~\ref{fig:fig3}a). With RF modulation, no added insertion loss was observed beyond fiber-to-chip coupling drift. There was a slight spectral broadening (7.1\% for the redshifted photons and 4.9\% for the blueshifted photons), mainly due to shift frequency fluctuation caused by timing jitters between the photon pulse and RF drive (see Methods). Since the optical pulse has a much smaller duration than the RF period, the magnitude of frequency shift is proportional to the ramp rate of the phase modulation and can be adjusted continuously by tuning the RF phase, following $K=-2\pi\times\pi\frac{V}{V_\pi}f_\mathrm{RF}\cos(\Delta\phi)$, where $f_\mathrm{RF}$ is RF frequency, and $\Delta\phi$ is the RF phase with respect to the optical pulses. Figure~\ref{fig:fig3}b shows the measured single-photon spectra with increasing $\Delta\phi$\, and their center wavelengths were extracted in Fig.~\ref{fig:fig3}c. The frequency shift underwent a sinusoidal change with a maximum range of $\pm$ 5.2 nm ($\pm$ 641 GHz). This shift frequency is about 3$\times$ higher than previous demonstrations using discrete, bulk LN modulators at visible wavelength ($\sim6\times$ higher when adjusted for wavelength since modulator $V_\pi$ decreases with wavelength). Equivalently, the RF power on the bulk modulator needs to be increased by 36$\times$ ($\sim$70 W based on Ref.~\cite{Wright2017}) to achieve the same shift frequency, which will likely exceed its power handling capability. Our result is also more than 4$\times$ higher than that based on optomechanical shearing~\cite{Fan2016}. Our shift frequency is currently limited by the frequency synthesizer and power amplifier instead of the device’s bandwidth or power handing threshold. Larger frequency shift can be achieved with even higher RF power and frequency, but more stable pump laser and better locking methods are needed to alleviate timing jitter induced bandwidth broadening~\cite{Karpinski2017}.

\begin{figure}
    \centering
    \includegraphics[width = 3in]{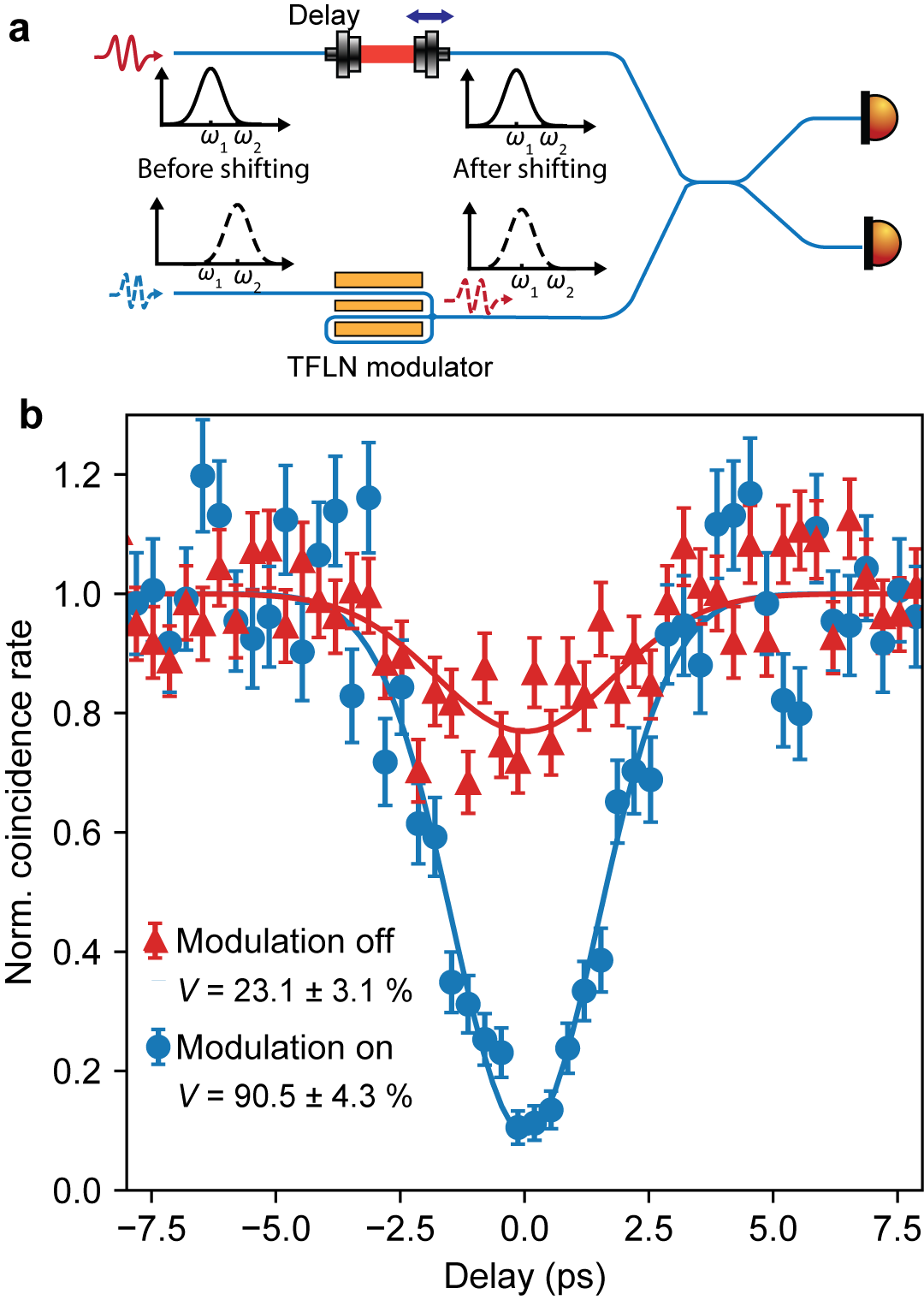}
    \caption{\textbf{Hong-Ou-Mandel (HOM) interference between two photons with different frequencies.} \textbf{a}, HOM interference set-up. The two input photons have FWHM bandwidth of 200 GHz and frequency difference of 154 GHz. The TFLN modulator was used to redshift the blue-detuned photon and erase the spectral distinguishability. The modulator was driven at 4.0 $V_\pi$ with a 13.6 GHz sinusoidal RF signal. \textbf{b}, Normalized coincidence rates after background subtraction as a function of path length delay. When the shearing was turned on, the interference visibility increased from 23.1 $\pm$ 3.1\% (due to partial spectral overlap) to 90.5 $\pm$ 4.3\%. Error bars indicate one standard deviation assuming Poisson counting statistics. Solid lines: Gaussian fitting.}
    \label{fig:fig4}
\end{figure}

To verify that the frequency shearing process does not introduce unintended modifications to the photons, we performed Hong-Ou-Mandel interference between twin photons with different frequencies. We first prepared nondegenerate photon pairs with a frequency detuning of 154 GHz by tuning the temperature of the nonlinear crystal (see Methods). Both signal and idler photons were filtered to a full-width half-maximum (FWHM) bandwidth of 200 GHz. The red photon went through a tunable delay, and the blue photon passed through the TFLN modulator. They then interfered at a beam splitter, and the coincidences at the two output ports were measured as a function of delay time (Fig.~\ref{fig:fig4}). High-visibility HOM interference requires the two photons to be indistinguishable. Without correcting their frequency differences, the two photons had a low interference visibility of 23.1 ± 3.1\% (background subtracted, see Methods), due to partial spectral overlap. By turning on the spectral shearing and redshifting the blue photon to erase the frequency distinguishability, we observed a high interference visibility of 90.5 $\pm$ 4.3\%. In this experiment, we reduced the RF drive on the modulator to 13.6 GHz and $\sim4.0V_\pi$ (i.e., $\sim$0.85 W RF power on the modulator), which optimized the tradeoff between intended frequency shift and temporal jitter-based spectral broadening as mentioned above. The high visibility two-photon interference between unshifted and shifted photons, which is a prerequisite for frequency-domain quantum computing and networking, indicates that shearing did not introduce unwanted distortion or noise to the photons.

\begin{figure*}[tbh!]
    \centering
    \includegraphics[width = 5.7in]{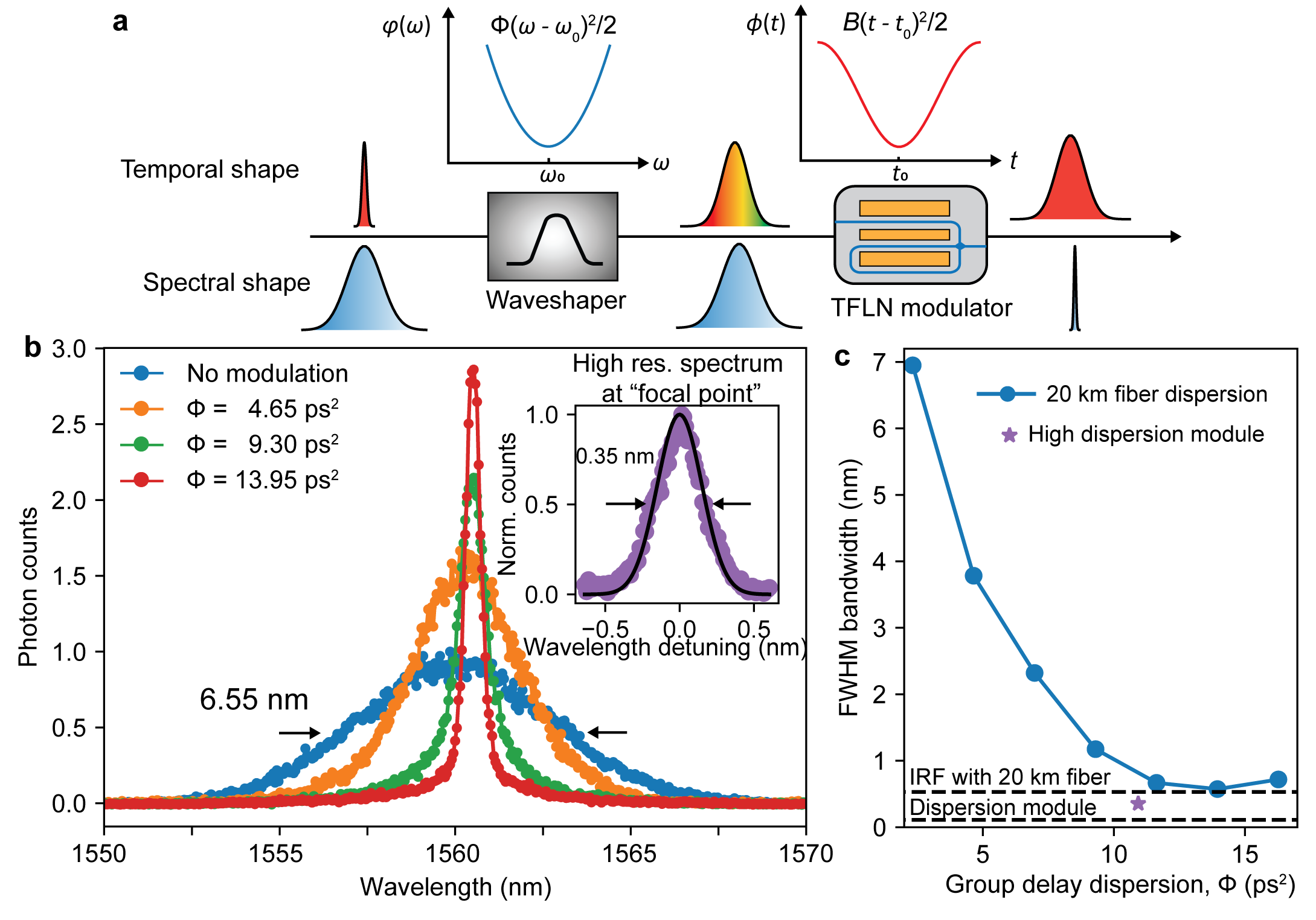}
    \caption{\textbf{Bandwidth compression of single photons.} \textbf{a}, Experimental implementation and illustration of the temporal and spectral evolution of the photon. A waveshaper first applies a quadratic spectral phase (or group delay dispersion, GDD) to the input photon, which stretches the wavepacket in time. The TFLN modulator then adds a quasi-quadratic temporal phase to the photon by synchronizing the photon arrival time to the valley of the sinusoidal phase modulation. This modulation acts as a time lens and compresses the photon bandwidth. \textbf{b}, Measured single-photon spectra with different GDDs. At the ``focal point'' ($\Phi = 1/B$), the 6.55 nm FWHM input photon (blue curve) was compressed to 0.35 nm (inset; black solid line: Gaussian fitting), corresponding to a compression factor of 18.7. \textbf{c}, Extracted FWHM bandwidths (from Gaussian fitting) of the output photons with different GDDs. Spectra in the main panel of (\textbf{b}) and blue circles in (\textbf{c}) were measured using 20 km fiber, which has an instrument response function (IRF)-limited resolution of 0.53 nm; inset of (\textbf{b}) and purple star in (\textbf{c}) were measured using a high dispersion module made of a chirped fiber Bragg grating, which has an IRF-limited resolution of 0.11 nm. The TFLN modulator was driven at 13.6 GHz and 4.0 $V_\pi$.}
    \label{fig:fig5}
\end{figure*}

Next, we configured the TFLN modulator as a time lens and demonstrated single-photon bandwidth compression. We first applied a quadratic spectral dispersion, $\Phi(\omega-\omega_0)^2/2$, to a nearly transform-limited single-photon pulse using a waveshaper, which introduced programmable phase delays at different wavelengths. We then directed the dispersed photons to the TFLN modulator and synchronized them to the valley of the sinusoidal phase modulation, imposing a nearly quadratic temporal phase of $\phi(t)=Bt^2/2$, with $B = 4\pi^3\frac{V}{V_\pi}f_\mathrm{RF}^2$ (see Fig.~\ref{fig:fig5}a). Intuitively, the spectral phase separates different frequency components in time, which then experience different amounts of frequency shearing and therefore lead to bandwidth compression. During the measurement, we kept the RF drive at 13.6 GHz and $\sim4.0V_\pi$, corresponding to $1/B\approx 10.9$ ps$^2$. The uncompressed input photon had a FWHM bandwidth of 6.55 nm (807 GHz). We gradually increased the group delay dispersion, $\Phi$, by re-programming the waveshaper. Doing so is equivalent to moving the ``object'' away from the lens and towards the ``focal point'' ($\Phi = 1/B$). As expected, the measured single-photon spectra and their FWHM bandwidths were compressed tighter and tighter with increasing $\Phi$, as shown in Figure~\ref{fig:fig5}b and c. Note that when $\Phi$ becomes too large, the dispersed pulse will start to overfill the time lens’ aperture ($\sim 1/f_\mathrm{RF}$), causing spectral distortion and spurious sidebands (see simulated spectra in Supplementary Figure~\ref{fig:fig_s3}). The slight frequency shifts in the measured spectra were likely caused by parasitic group delays added by the waveshape at different spectral phase settings, which caused optical pulses to move away from the valley of the RF drive. We also observed spectral broadening after engaging the modulation (from 6.55 nm to 6.95 nm, without adding spectral dispersion), which was due to the timing jitter between the RF drive and optical pulse. Here, the spectral resolution was 0.53 nm (determined by the timing jitter of the SNSPD and dispersion given by the 20 km single-mode fiber), which artificially broadened the measured spectrum. To uncover the actual bandwidth of the compressed photons, we performed high-resolution spectroscopy at $\Phi = 1/B$ using a Bragg grating based dispersion module, achieving a spectral resolution of $\sim$0.11 nm~\cite{Chen2017}. The measured FWHM bandwidth was 0.35 nm (43.1 GHz), corresponding to a compression factor of 18.7 (inset of Fig.~\ref{fig:fig5}b and star in Fig.~\ref{fig:fig5}c), about 3$\times$ what was demonstrated using discrete bulk LN modulators at visible wavelengths~\cite{Karpinski2017}. 

In summary, we have demonstrated high-performance electro-optic single-photon spectral shearing and bandwidth compression using an integrated TFLN phase modulator. Besides outstanding performance, a major advantage of our approach is its integrability with other essential components on the TFLN platform~\cite{Zhu2021,Saravi2021}, such as sources~\cite{Javid2021Ultra-broadbandChip,Aghaeimeibodi2018}, detectors~\cite{Lomonte2021Single-photonCircuits}, memories~\cite{Dutta2021}, and a complete set of linear optical components~\cite{Zhu2021}, for realizing complex photonic circuits and functionalities. By cascading multiple modulators and dispersion units, which may be implemented using dispersion-engineered waveguides or integrated wavelength division multiplexer (WDM) followed by a bank of phase shifters, one may effectively replace the bulk waveshaper and achieve arbitrary spectral control of light fully on-chip. Our work accelerates the development of optical quantum processors and networks, which require low-loss control, and high-visibility interference, of individual photons in the frequency domain to yield an advantage. New functionalities are expected beyond mode matching, for instance, the time lens increases the duration of a photon wavepacket far beyond the temporal resolution of a single-photon detector~\cite{Korzh2020a}, which could allow interference of widely detuned photons~\cite{Zhao2014EntanglingForward}, while the shifter, in combination with a WDM and several such detectors, could allow circumventing rate-limiting detector recovery times. Precise control of the phase of light can also open up new possibilities for high-dimensional quantum information~\cite{Brecht2015PhotonScience}. Our approach is scalable, does not require optical pump or filtering, and can be extended to a wide range of operating wavelengths. We expect it to become a useful building block for realizing temporal-spectral control of light in both quantum and classical applications, such as frequency-encoded communication and computation, and ultrafast pulse generation and measurement.

\textbf{Acknowledgements.}
We thank Brian J. Smith, Karl Berggren, Marco Colangelo, Marco Turchetti, and Mina Bionta for helpful discussions and assistance in measurement. This work is supported by Harvard Quantum Initiative (HQI), ARO/DARPA (W911NF2010248), AFOSR (FA9550-20-1-01015), DARPA LUMOS (HR0011-20-C-0137), DOE (DE-SC0020376), NSF (EEC-1941583), and AFRL (FA9550-21-1-0056). D.Z. acknowledges support by HQI post-doctoral fellowship and A*STAR SERC Central Research Fund (CRF). N.S. acknowledges support by the Natural Sciences and Engineering Research Council of Canada (NSERC). Device fabrication was performed at the Harvard University Center for Nanoscale Systems.

\textbf{Author contributions.}
D.Z., M.Yu, N.S., and M.L. conceived and designed the experiment. M.Yu designed the modulator. L.H., C.R., and M.Z. fabricated the modulator. D.Z., C.C., M.Yu, and L.S. carried out the measurement and analyzed the data with the help of Y.H., C.J.X, M.Yeh, S.G. and N.S. All authors contributed to writing the manuscript. M.L. and F.N.C.W supervised the project.

\textbf{Competing interests.}
M.Z., L.H., C.R., and M.L. are involved in developing lithium niobate technologies at HyperLight Corporation.

\textbf{Disclaimer.}
The views, opinions and/or findings expressed are those of the author and should not be interpreted as representing the official views or policies of the Department of Defense or the U.S. Government.

\section*{Methods}
\paragraph{Modulator characterization}
The half-wave voltage and bandwidth of the modulator were characterized using a telecom continuous-wave laser. The modulator was driven by a signal generator with calibrated output power. The RF drive was swept from 10 to 40 GHz with 100 MHz steps at two different power levels (13.0 dBm and 19.0 dBm). The output optical signal was captured by an optical spectrum analyzer, and the $V_\pi$ was estimated by fitting the sideband powers using a Bessel function, $|J_n (\pi\frac{V}{V_\pi})|^2$, where n is the sideband number, V is the drive voltage. The measured $V_\pi$ as a function of frequency is shown in Supplementary Fig.~\ref{fig:fig_s1}b. Light was coupled in and out of the modulator chip using lensed fibers. The total insertion loss was 11 dB, including fiber-to-chip coupling loss of $\sim$4.5 dB per facet. The coupling and on-chip loss could be reduced to as low as 0.5 dB/facet and 2.7 dB/m by optimized design~\cite{Ying2021} and fabrication~\cite{Zhang2017}. No increase in insertion loss was observed when turning on RF modulation. More details about the double-pass modulator can be found in Ref.~\cite{Yu2021}.

\paragraph{Locking between photon pulse and RF drive}

The pump laser for the SPDC photon generation has a repetition rate of $\sim$80 MHz. Its electrical synchronization signal, produced by sampling the laser output with a fast photodetector, was filtered by a bandpass filter around 160 MHz. The filtered signal was then used as a frequency reference for a frequency synthesizer (Analog Devices EV-ADF4371SD2Z) to generate the phase-locked RF drive for the TFLN modulator. The synthesizer did not have a calibrated output power from the manufacturer. To estimate the RF power on the modulator (after power amplifiers, attenuators, and cables), we sent a classical continuous-wave laser into the modulator and measured the generated sideband using an optical spectrum analyzer. The drive voltage was fitted from the sideband distribution (see Supplementary Fig.~\ref{fig:fig_s2}) and calculated using the calibrated modulator $V_\pi$ (Supplementary Fig.~\ref{fig:fig_s1}b). Due to the drift of laser repetition rate and limited response time of the phase-locked oscillator, there existed timing jitters between the photon pulses and RF drive. This jitter was measured to be $\sim$15 ps FWHM at $\sim$5 GHz and increased with increasing frequency. Direct measurements at 13.6 GHz and 27.5 GHz were not performed due to the limited bandwidth of our oscilloscope. The timing jitter would induce spectral broadening due to the fluctuation in the amount of frequency shifts. This phenomenon became more prominent when the relative RF phase $\Delta\phi$ deviated from integer multiples of $\pi$ (i.e., fastest rising/falling slopes), as observed in Fig.~\ref{fig:fig2}b, which is consistent with previous studies and can be alleviated with a more stable pump laser or better locking method~\cite{Karpinski2017}. 

\paragraph{Single-photon spectrum measurement.}
To measure single-photon spectra, the photons were sent through a 20 km fiber. The fiber dispersion performs frequency-to-time conversion (333.8 ps/nm) and was used to reconstruct the spectrum from the measured delay between the photon arrival time and laser trigger~\cite{Chen2017,Avenhaus2009}. The pump laser had a repetition rate of 80 MHz, and the SNSPDs had a FWHM timing jitter $\sim$180 ps, giving a spectral resolution of 0.53 nm and range of 37 nm. In the time lens experiment (Fig.~\ref{fig:fig5}), a fiber Bragg grating based dispersion unit (1.88 ns/nm) was used for high resolution spectrum measurement, which gave a resolution of 0.11 nm and range of 6.6 nm.  

\paragraph{Hong-Ou-Mandel interference.}
In the HOM interference measurement, we tuned the temperature of the crystal to 48.5$^\circ$C and set the two waveshapers as 200\,GHz FWHM spectral filters with frequency detuning of 154 GHz. This separation was partly limited by the temperature response of the nonlinear crystal, which was originally designed for degenerate operation. The coincidence rates between the two SNSPDs were monitored as a function of delay (100\,$\upmu$m/0.33 ps steps). The raw coincidence rate far away from the interference dip was 10.94 cps, and the background coincidences (measured by blocking individual beam path) for the airgap and device paths were 1.88 cps and 0.03 cps, respectively. The rate was mainly limited by the narrow filter bandwidth and insertion loss of the two waveshapes, and the background coincidences were likely due to multipair events and polarization misalignment. The acquisition time for each data point was 15 s. The measured coincidence curve was fitted using a Gaussian function, and the visibility was calculated as $(N_\mathrm{max}-N_\mathrm{min})/N_\mathrm{max}$, where $N_\mathrm{max/min}$ are maximum/minimum coincidence rates. In the main text, we presented background-subtracted HOM interference visibility of 23.1 $\pm$ 3.1 \% and 90.5 $\pm$ 4.3 \% for the unshifted and shifted cases, respectively. The non-zero visibility for the unshifted case was due to the partial spectral overlap of the two photons. Without background subtraction, the raw values were 19.2 $\pm$ 2.6 \% and 74.6 $\pm$ 3.6\%. As a reference, we set the passband of the two waveshapers to be the same while keeping the crystal temperature at 48.5$^{\circ}$C, and measured HOM interference visibility (without frequency shifting) to be 86.2 $\pm$ 3.2\% with background subtraction and 73.3 $\pm$ 2.8\% without background subtraction. This shows that the maximum visibility was not limited by the shearing process, but mainly by the measurement setup, such as the resolution and accuracy of the spectral filters and nonoptimal spectral shape of the source at this operating temperature.

\beginsupplement
\widetext
\clearpage
\begin{center}
	\textbf{\large Supplementary Figures}
\end{center}

\begin{figure*}[tbh!]
    \centering
    \includegraphics[width = 3in]{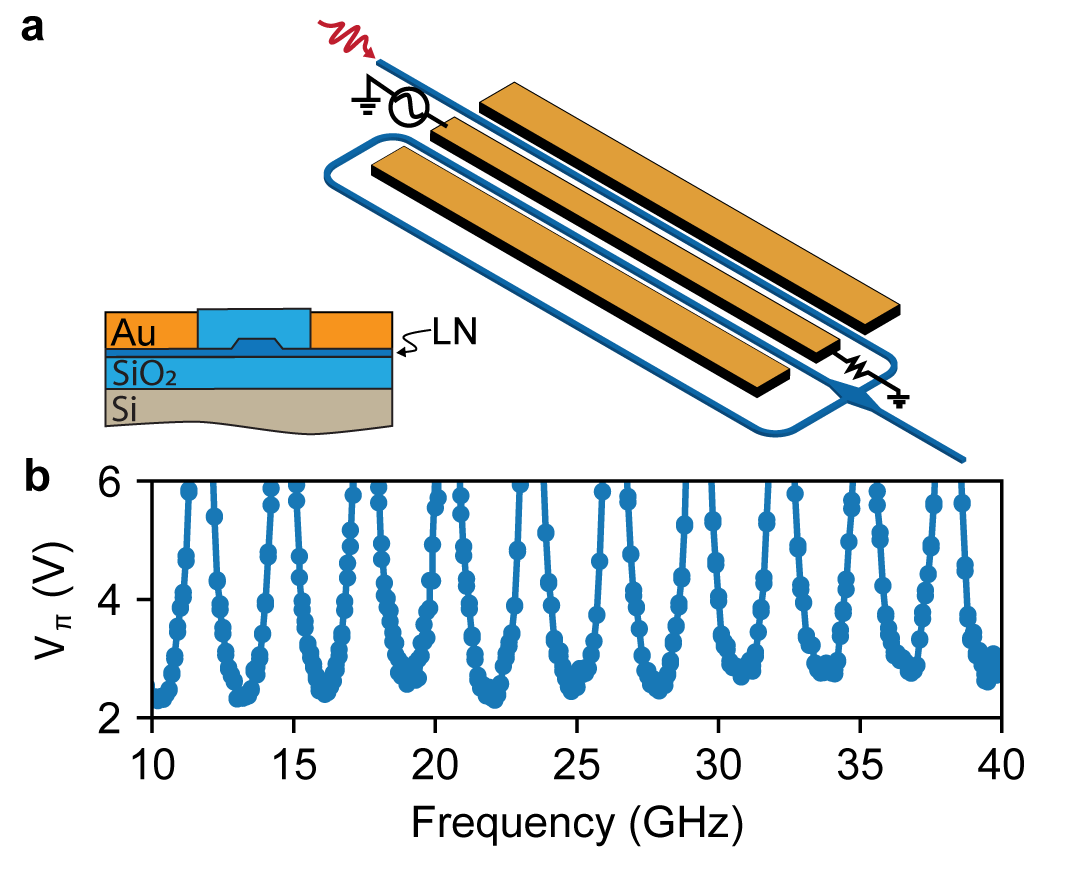}
    \caption{\textbf{Device geometry and performance.} \textbf{a}, Device layout and cross-section of the TFLN electro-optic phase modulator. The modulator adopts a double-pass configuration. This special design increases the interaction length between optical and RF fields, resulting in a reduced $V_\pi$. The physical size of the modulator (active area) is 2 cm $\times$ 600 $\upmu$m. \textbf{b}, Measured $V_\pi$ at 1550 nm wavelength. Efficient modulation happens when light is in-phase with the RF drive when it re-enters the coplanar waveguide electrode. At these phase-matched frequencies, the measured $V_\pi$ is between 2.3-2.8 V from 10 to 40 GHz.}
    \label{fig:fig_s1}
\end{figure*}

\begin{figure*}[tbh!]
    \centering
    \includegraphics[width = 4.5in]{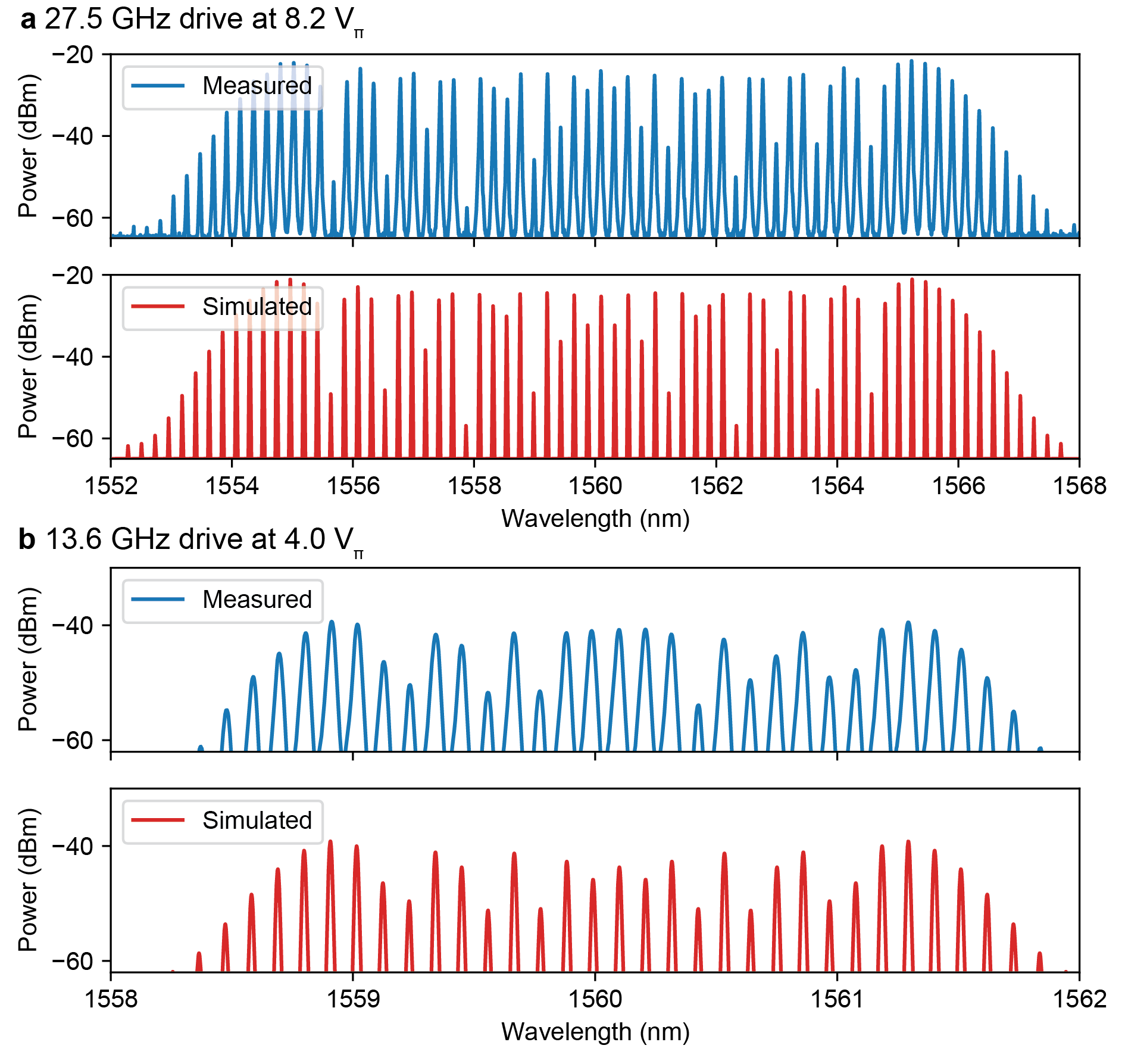}
    \caption{\textbf{Modulation sidebands when the TFLN modulator was driven at conditions used in the main text.} A classical continuous-wave laser was injected to the modulator and the output was measured using an optical spectrum analyzer. The driving amplitude in terms of number of $V_\pi$ was estimated by fitting the measured spectra using a Bessel function. \textbf{a}, Driving condition in Fig.~\ref{fig:fig3} in the main text. \textbf{b}, Driving condition in Fig.~\ref{fig:fig5} and \ref{fig:fig5} in the main text.}
    \label{fig:fig_s2}
\end{figure*}

\begin{figure*}[tbh!]
    \centering
    \includegraphics[width = 4in]{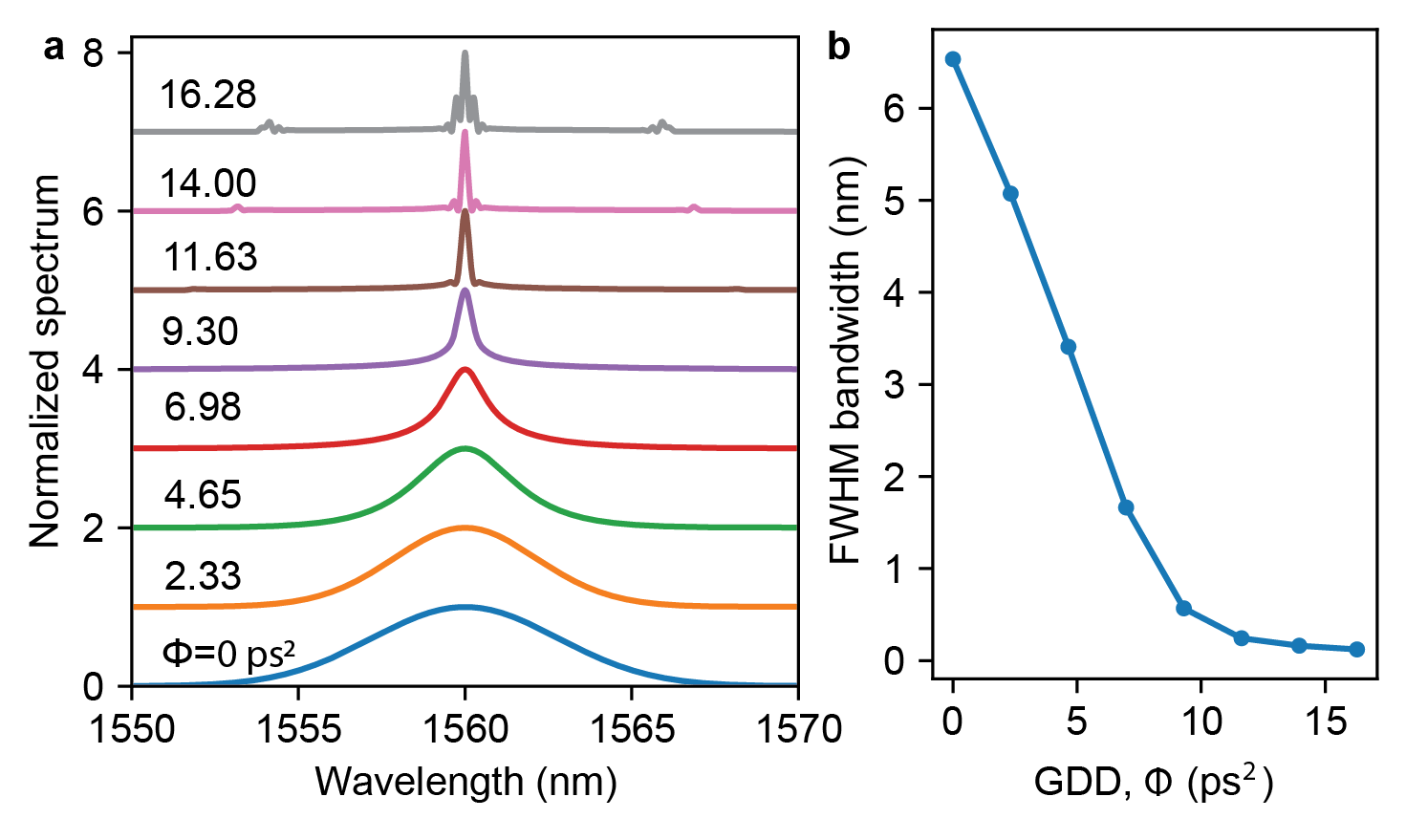}
    \caption{\textbf{Time lens simulation.} Simulated spectral shape (\textbf{a}) and FWHM bandwidth (\textbf{b}) after time lensing with increasing group delay dispersion (GDD) applied. When GDD exceeds $\sim$10 ps$^2$, the dispersed optical pulse that enters the time lens become too wide and starts to overfill the time lens' effective aperture. As a result, spectral distortion as well as sidebands are generated. However, these fine features are difficult to resolve in experiments due to limited measurement resolution. Instead, they manifest as an effectively broadened spectrum, as shown in Fig.~\ref{fig:fig5}c in the main text. In this simulation, the input photon has a center wavelength of 1560 nm and FWHM bandwidth of 6.55 nm. RF modulation is at 13.6 GHz and 4.0 $V_\pi$. }
    \label{fig:fig_s3}
\end{figure*}

\end{document}